\documentclass{ifacconf}
\newcommand{\R}{\mathbb{{R}}}
\newcommand{\nats}{\mathbb{{N}}_{\geq 0}}
\newcommand{\He}{\operatorname{He}}
\newcommand{\dom}{\operatorname{dom}}
\newcommand{\rank}{\operatorname{rank}}
\newcommand{\prj}{\operatorname{prj}}
\newcommand{\xSpace}{\mathcal{X}}
\newcommand{\np}{n_p}
\newcommand{\nv}{n_u}
\newcommand{\nc}{n_c}
\newcommand{\xp}{x_p}
\newcommand{\xpdot}{\dot{x}_p}

\newcommand{\xc}{\xi}
\newcommand{\up}{u_p}
\newcommand{\range}{\operatorname{Im}}

\newcommand{\QEDA}{$\hfill\blacksquare$}
\usepackage{amssymb,mathtools}
\usepackage{amsmath}
\usepackage{scalerel,balance}
\usepackage{graphicx}      
\usepackage{natbib}  
\usepackage{psfrag}
\usepackage{booktabs}   
\usepackage{siunitx}    
\usepackage{xcolor,comment}
\newtheorem{assumption}{Assumption}
\newtheorem{problem}{Problem}
\newtheorem{lemma}{Lemma}
\newtheorem{proposition}{Proposition}
\newtheorem{theorem}{Theorem}

\newtheorem{remark}{Remark}
\newtheorem{definition}{Definition}
\begin{document}
\begin{frontmatter}

\title{Robust Aperiodic Sampled-Data Washout Control for Uncertain Affine Systems} 

\thanks[footnoteinfo]{This research was funded by the University of Perugia, Fondo progetti di ateneo WP 4.3, progetto COLONIZE.}

\author{F. Giorgetti,} 
\author{F. Crocetti,} 
\author{M. L. Fravolini,}
\author{F. Ferrante}

\address{Dipartimento d'Ingegneria, Università degli Studi di Perugia (e-mails: folco.giorgetti@dottorandi.unipg.it, \{francesco.crocetti, francesco.ferrante, mario.fravolini\}@unipg.it.}

\begin{abstract}               
In this paper, we address the problem of designing an aperiodic sampled-data controller stabilizing the zero-input equilibrium of an uncertain affine plant. The closed-loop system is modeled as a hybrid dynamical system incorporating a timer triggering the occurrence of the sampling events and two memory states storing the value of the controller state and controller output at each sampling time. Necessary and sufficient conditions on the controller parameters are given to establish the sought property. A constructive controller design algorithm based on sum-of-squares programming is given. A numerical example illustrates the effectiveness of the approach.  
\end{abstract}


\end{frontmatter}

\section{Introduction}
\subsection{Motivation and Background}
The unforced equilibrium of a system refers to the state where the system variables do not change over time in the absence of external input or disturbances. In practice, determining the exact equilibrium can be surprisingly challenging when facing complex (possibly nonlinear) dynamics \cite{https://doi.org/10.1002/rnc.1819} or uncertain operating conditions. Such difficulties often lead to an uncertain equilibrium scenario where the nominal equilibrium is inaccurately known or drifts with external disturbances.
A range of approaches have been proposed to tackle the stabilization of systems with unknown or uncertain equilibrium points. First results addressing the problem of the stabilization of chaotic systems with unknown equilibrium states using delayed feedback control can be traced back in \cite{PYRAGAS1992421} and are followed by \cite{ 975474, hovel2005control}; just to mention a few. In \cite{7082644, huijberts2006linear}, the authors proposed an adaptive control scheme in which the uncertain parameter is the operating point. The most recent approaches involve state-derivative feedback controllers \cite{shigekuni2013stabilization, arthur2020robust, 9483104} and washout filter approaches \cite{1383925, takimoto2007reduced}. Among these methods, washout filters and derivative feedback stand out for their wide applicability. Derivative feedback control has been proposed for systems where it is feasible or even easier to measure state-derivative signals than the states themselves. For example, accelerometers and inertia measurement units provide practical solutions for state derivative measurements in translational and rotational systems \cite{6212314, doi:10.1177/1077546317711335}. However, the design of derivative feedback parameters can be intricate, and practitioners must contend with issues like sensor noise amplification and ensuring that the closed-loop system’s poles lie at desired locations. Washout filters are high pass filters that ``wash out'' steady state inputs, while passing transient inputs. The main benefit of using washout filters is that all the equilibrium points of the open-loop system are preserved. A common challenge lies in selecting appropriate filter parameters, as poor tuning can compromise performance or stability. In this work, we build on the central ideas of washout control to tackle the stability problem of sampled-data systems subject to aperiodic sampling and unknown unforced equilibria. Recent years have witnessed extensive research on the stability of sampled-data systems, with a comprehensive overview provided in \cite{HETEL2017309}. 
\subsection{Contributions}
In this paper, we address the problem of aperiodic sampled-data washout control design for uncertain affine plants. In particular, we consider a scenario in which the plant dynamics and input matrices are subject to norm bounded uncertainties and  drift term is unknown. Due to uncertainties in the plants dynamics and aperiodic sampling, the typical approach consisting of relying on a discretized model of the plant is not viable. To overcome this drawback, we adopt the hybrid systems framework introduced in \cite{Goebel2012}. Within this setting, we provide sufficient and necessary conditions on the controller parameters to ensure global exponential stability of a compact set containing the unknown unforced equilibrium of the plant. Based on these conditions and by relying on a similar approach as in \cite{BRIAT20133449}, sufficient conditions in the form of clock-dependent matrix inequalities are established. These conditions are then used to devise a computationally affordable control design algorithm based on sum-of-squares programming. It is worth to stress that, while the application of washout filters for discrete-time systems is studied in \cite{takimoto2007reduced}, to the best of our knowledge, the design of washout filters for aperiodic sampled-data systems with uncertain parameters has not been addressed. The remainder of the paper is organized as follows. Section~\ref{sec:ProblemSetting} introduces the problem we solve, some standing assumptions, and a description of the hybrid model of the closed-loop system. Section~\ref{sec:MainResults} presents necessary and sufficient conditions for the solution of the considered problem. Section~\ref{sec:ControllerDesign} proposes a constructive approach for controller design. 
Section~\ref{sec:Examples} showcases the effectiveness of our approach in a numerical example. 

\subsection{Notation}
The symbols $\R$, $\nats$, $\R^n$, $\R^{n\times m}$, and $\mathbb{S}^{n}$ represent, respectively, the set of the real numbers, natural numbers (including zero), the $n$-dimensional Euclidean space, the set of $n \times m$ real matrices, and the set of $n\times n$ symmetric matrices. The symbol $x^\top$ ($A^\top$) denotes the transpose of the vector $x$ (matrix $A$). For a square matrix $A$, we use the notation $\He(A)\coloneqq A+A^\top$. The notations $\range(M)$ and $\rank(M)$ indicate, respectively, the image and rank of the matrix $M$. The symbol $\star$ stands for a symmetric block in a symmetric matrix. Given a symmetric matrix $A$, $A\prec0$ ($A\succ0$) means that $A$ is negative (positive) definite. We use the notation $(x,y)=\begin{bmatrix}
    x^\top&y^\top
\end{bmatrix}^\top$. Given a function $x\colon\R\to\R^n$, the notation $x(t^-)$ indicates $\displaystyle\lim_{s\to t^-}x(t)$, when this limit exists. The symbol $\dom f$ stands for the domain of the function $f$.
\subsection{Preliminaries on hybrid systems}
In this paper we consider hybrid  dynamical systems in the framework \cite{Goebel2012} represented as:
\begin{equation}
\label{eq:HybridPlant}
\mathcal{H}\left\{
\begin{array}{lcll}
\dot{x}&=&f(x),&\quad x\in \mathcal C,\\
x^+&=&g(x),&\quad x\in  \mathcal D.
\end{array}\right.
\end{equation}
where $x\in\R^{n}$ is the state vector, $f\colon\R^{n}\rightarrow\R^{n}$ denotes the flow map and $g\colon\R^{n}\rightarrow\R^{n}$ the jump map, while the sets $\mathcal C\subset\R^{n}$ and $\mathcal D\subset\R^{n}$ refer to the flow and the jump sets, respectively. A set $\mathcal{E}\subset\R_{\geq 0}\times \mathbb{N}_{\geq 0}$ is a \emph{hybrid time domain} (\emph{HTD}) if it is the union of a finite or infinite sequence of intervals $[t_j, t_{j+1}]\times\{j\}$, with the last interval (if existent) of the form $[t_j,T)$ with $T$ finite or $T=\infty$. A function $\phi\colon\dom\phi\rightarrow\R^n$ is a hybrid arc if $\dom\phi$ is a HTD and $t\mapsto\phi(t, j)$ is locally absolutely continuous for each $j$. A solution to \eqref{eq:HybridPlant} is any hybrid arc that satisfies its dynamics. A solution $\phi$ to $\mathcal{H}$ is maximal if its domain cannot be extended and it is complete if its domain is unbounded. Given a set $\mathcal M$, we denote by $\mathcal{S}_{\mathcal{H}}(\mathcal M)$ the set of all maximal solutions $\phi$ to $\mathcal{H}$ with $\phi(0,0)\in \mathcal M$. If no set $\mathcal{M}$ is mentioned, $\mathcal{S}_{\mathcal{H}}$ is the set of all maximal solutions to $\mathcal{H}$. We say that $\mathcal{H}$ satisfies the hybrid basic conditions if: $\mathcal{C}$ and $\mathcal{D}$ are closed and $f$ and $g$ are continuous. The following definitions are used throughout the paper. 
\begin{definition}
Let $\mathcal{A}\subset\R^n$ be compact. We say that $\mathcal{A}$ is globally exponentially stable (GES) for \eqref{eq:HybridPlant} if any maximal solution to \eqref{eq:HybridPlant} is complete and there exist positive scalars $\gamma, \kappa$ such that for any $\phi\in\mathcal{S}_{\mathcal{H}}$
$$
\vert \phi(t,j)\vert_{\mathcal{A}}\leq \kappa e^{-\gamma (t+j)}\vert \phi(0,0)\vert_{\mathcal{A}}\quad\forall (t, j)\in\dom\phi.
$$\hfill$\triangle$
\end{definition} 
\begin{definition}
Let $\phi$ be a hybrid arc. The set $\Omega(\phi)$ corresponds to the set of points $x\in\R^n$ for which there exists a sequence $\{(t_i, j_i)\})\subset\dom\phi$ such that $\lim_{i\to\infty} t_i+j_i=\infty$ and $$\lim_{i\to\infty} \phi(t_i, j_i)=x.$$
\end{definition}

For more details on hybrid systems, the reader is referred to \cite{Goebel2012}.
\section{Problem Statement}
\label{sec:ProblemSetting}
\subsection{Problem setting}
We consider the following class of uncertain affine plants  
\begin{equation}
\label{eq:plant}
\begin{aligned}
    &\dot{x}_p= A\xp+Bu_p+d
\end{aligned}
\end{equation}
where $\xp\in\mathbb{R}^{n_p}$ is the plant state, $u_p\in\R^{n_u}$ is the control input, and $d\in\R^{n_p}$ is an unknown nonzero constant vector. Matrices $A$ and $B$ are real and unknown. We assume that, for all $d\in\R^{n_p}$, \eqref{eq:plant} admits a unique unforced equilibrium point. Our goal is to design a feedback controller asymptotically driving the plant towards such an equilibrium without any knowledge of the value of $d$. We consider a scenario in which plant \eqref{eq:plant} is controlled via the following sampled-data dynamic state-feedback controller
\begin{equation}
\label{eq:Controller}
\begin{array}{llc}
&\xi(t_k)=L\xi(t_k^-)+Gu_c(t_k^-)\\[2mm]
&y_c(t_k)=K\xi(t_k^{-})+Ru_c(t_k^{-})
\end{array}
\end{equation}
where $\xc$, $u_c$ and $y_c$, denote, respectively, the controller state, input, and output.
The sequence $t_1, t_2,\dots$ denote the sampling instances and $L$, $G$, $K$, and $R$ are the controller parameters to be designed. We assume plant \eqref{eq:plant} is equipped with a zero-holder-device holding the control input constant in between samples. In particular, by setting $\up=y_c$ and $u_c=\xp$, the closed-loop system reads
\begin{equation}
\label{eq:CL}
\begin{aligned}
&\xpdot(t)= A\xp(t)+B y_c(t_k)+d&\forall t\in[t_k, t_{k+1})\\
&\xi(t_k)=L\xi(t_{k}^-)+G\xp(t_k^-).
\end{aligned}
\end{equation}
The following standing assumptions are considered through the paper.
\begin{assumption}
\label{assu:sampling}
There exist $0<T_1\leq T_2$ such that the following holds
$$
\begin{aligned}
&0\leq t_1\leq T_2\\
&T_1\leq t_{k+1}-t_{k}\leq T_2&\forall k\in\mathbb{N}_{>0}.
\end{aligned}
$$
\hfill$\triangle$
\end{assumption}
\medskip

\begin{assumption}
\label{assu:Bcol}
Matrix $B$ is full-column rank.
\hfill$\triangle$
\end{assumption}
\medskip

\begin{assumption}
\label{assu: X0}
For all $d\in \R^{n_p}$, the set
$$
\mathcal{X}_d\coloneqq\{\xp\in\R^{\np}\colon A\xp+d=0\}
$$
is nonempty and is a singleton. \hfill$\triangle$
\end{assumption}
\medskip

\begin{assumption}
\label{assu: Incertezza}
There exist $A_0\in \R^{\np\times\np}$, $B_0\in\R^{\np\times\nv}$, $E_{\Delta}\in\R^{m_{\Delta}\times\np}$, $F_{\Delta}\in\R^{m_{\Delta}\times\nv}$, and $D_{\Delta}\in\R^{\np\times n_{\Delta}}$ such that 
$$
A=A_0+D_{\Delta}\Delta E_{\Delta},\quad B=B_0+D_{\Delta}\Delta F_{\Delta} 
$$
for some
$$
\Delta\in\mathcal{U}\coloneqq\left\{\Delta\in\R^{n_{\Delta}\times m_{\Delta}}\colon \Delta^\top \Delta\preceq I\right\}.
$$
\hfill$\triangle$
\end{assumption}
Assumption~\ref{assu:sampling} ensures that the sequence of sampling instances $\{t_k\}$ is monotonically increasing and does not contain any accumulation point. In particular, $T_1$ and $T_2$, represent, respectively the lower and upper bound on the distance between two consecutive samplings. When $T_1=T_2$, sampling occurs periodically. Assumption~\ref{assu:Bcol} is not restrictive and can be always fulfilled by removing linear dependent columns. Assumption \ref{assu: X0} implies that $A$ is nonsingular and guarantees the existence of a unique unforced equilibrium point for any $d\in\R^{\np}$. Assumption~\ref{assu: Incertezza} is fairly standard in the context of robust control. Alternative parameterizations of the uncertainty can be used. 
\subsection{Hybrid modeling}
Due to the interplay of sampling events and continuous evolution of the plant, we model the interconnection of plant \eqref{eq:plant} and controller \eqref{eq:Controller} as a hybrid dynamical system. More in particular, we rely on the framework in \cite{Goebel2012}. To this end, we consider a clock variable $\tau\in [0, T_2]$ triggering the sampling events. In particular, we let the variable $\tau$ flow as long as $\tau\in [0, T_2]$ and trigger a jump when $\tau\in[T_1, T_2]$. At jumps, the value of $\tau$ is reset to zero. In addition, we introduce two variables $q\in\R^{m}$ and $\xc\in\R^{n_c}$ to store, respectively, the values of the controller state and controller output at the sampling times. By taking $x\coloneqq(z,\tau)\coloneqq(\xp, \xc, q,\tau)\in\xSpace\coloneqq\R^{n+1}$, where $n\coloneqq n_p+n_u+n_c$, the closed-loop system can be modeled by the following hybrid system $\mathcal{H}$ with state $x$
\begin{equation}
\mathcal{H}\left\{ \begin{array}{lll}
&\begin{cases}
\dot{z}=\mathrm{F} z+\mathrm{B}_f d\\
\dot{\tau}=1
\end{cases}&\mathcal{C}\coloneqq\{x\in\xSpace\colon \tau\in[0, T_2]\}\\
&\begin{cases}
z^+=\mathrm{J}z\\
\tau^+=0
\end{cases}&\mathcal{D}\coloneqq\{x\in\xSpace\colon \tau\in[T_1, T_2]\}\\
 \end{array}  
 \label{eq:HCL}\right.
\end{equation}
where 
$$
\mathrm{F}\coloneqq\begin{bmatrix}
  A&0&B\\
0&0&0\\
0&0&0
\end{bmatrix}, \mathrm{B}_f\coloneqq\begin{bmatrix}
    I\\0\\0
\end{bmatrix}, \mathrm{J}\coloneqq\begin{bmatrix}
I&0&0\\
G&L&0\\
R&K&0
\end{bmatrix}.
$$
Next we outline some structural, yet elementary, properties for hybrid system \eqref{eq:HCL}.
\begin{fact}
\label{fact:HybridBasic}
The following properties hold:
\begin{itemize}
    \item[$(a)$] Hybrid system \eqref{eq:HCL} satisfies the hybrid basic conditions;
    \item[$(b)$] Any maximal solution to \eqref{eq:HCL} is complete.
\end{itemize}   
\end{fact}
We are now in a position to formally state the problem we solve in this section. 
\begin{problem}
\label{prob:state_feedback}
Design $L$, $G$, $K$, and $R$, such that for all $d\in\R^{\np}$, there exists a nonempty compact set 
$$
\mathcal{A}_d\subset\mathcal{X}_d\times\R^{\nc}\times\{0\}\times [0, T_2]
$$ 
that is GES for system \eqref{eq:HCL}.
\end{problem}
Before stating the main results we state the following fact
\begin{fact}\label{fact:A_nonsingular}
    Problem~\ref{prob:state_feedback} is solvable only if matrix $A$ is nonsingular\footnote{Fact~\ref{fact:A_nonsingular} highlights that Assumption~\ref{assu: X0} is necessary for the solvability of Problem~\ref{prob:state_feedback}}.
\end{fact}

\section{Main Results}
\label{sec:MainResults}
Next we state a necessary condition for the solution to Problem~\ref{prob:state_feedback}. This condition relies on the following ancillary result, whose proof can be found in the appendix.
\begin{lemma}
Let $d\in\R^{\np}$. Suppose that $(L,G,K, R)$ solves Problem~\ref{prob:state_feedback}. If the pair $(L, K)$ is observable, then there exists $\bar{\xc}\in\R^{\nc}$ such that for all $\phi\in\mathcal{S}_{\mathcal{H}}$  
$$
\prj_{z}(\Omega(\phi))=\{-A^{-1}d, \bar{\xc}, 0\}
$$
where, for all $x\in\mathcal{C}$
$$\prj_{z}(x)\coloneqq\{z\in\R^{n}\colon\exists\tau\in[0, T_2]\,\,\text{s.t.}\,\,x=(z,\tau)\}.$$
\label{lemm:singleton}
\hfill$\triangle$
\end{lemma}
\medskip

\begin{proposition}
Assume that the pair $(L, K)$ is observable. Then, $(L, G, K, R)$ solves Problem~\ref{prob:state_feedback} only if
\begin{equation}
\rank\begin{bmatrix}
R&K\\
G&L-I
\end{bmatrix}=\rank\begin{bmatrix}
K\\
L-I
\end{bmatrix}.
    \label{eq:Z0}
\end{equation}
\label{prop:nec}
\end{proposition}
\begin{pf}
Let $d\in\R^{n_p}$. From Lemma~\ref{lemm:singleton}, there exists 
$\bar{\xc}\in\R^{\nc}$ such that for all $\phi\in\mathcal{S}_\mathcal{H}$  
$$
\prj_{z}(\Omega(\phi))=\{-A^{-1}d, \bar{\xc}, 0\}.
$$
Now recall that $\Omega(\phi)$ is invariant for \eqref{eq:HCL}. Let $\psi = (\psi_{\xp}, \psi_{\xc}, \psi_{q},\psi_{\tau})\in\mathcal{S}_{\mathcal{H}}(\Omega(\phi))$ and $0=t_0\leq t_1\leq t_2\leq t_{j}\leq\dots$ be such that
$$
\dom\psi=\bigcup_{j=0}^\infty [t_{j}, t_{j+1}]\times\{j\}.
$$
From the flow dynamics, for all $j\in\nats$, one has that
\begin{equation}
\psi_{q}(t, j)=0, \quad\psi_{\xc}(t, j)=\bar{\xi}\quad \forall t\in[t_{j}, t_{j+1}]
\label{eq:flow_id}
\end{equation}
while from the jump dynamics
\begin{equation}
\begin{aligned} 
\bar{\xc}=\psi_{\xc}(t, j+1)=L\psi_{\xc}(t, j)-GA^{-1}d\\
0=\psi_{q}(t, j+1)=K\psi_{\xc}(t, j)-RA^{-1}d.
\end{aligned}
\label{eq:jump_id}
\end{equation}
By combinining \eqref{eq:flow_id}-\eqref{eq:jump_id}, one gets
$$
\begin{aligned} 
(L-I)\bar{\xc}=GA^{-1}d\\
K\bar{\xc}=RA^{-1}d.
\end{aligned}
$$
Then, since $d$ is arbitrary, the latter implies 
$$
\range\begin{bmatrix}
  G\\
  R
\end{bmatrix}\subset\range\begin{bmatrix}
  L-I\\
  K
\end{bmatrix}
$$
which is equivalent to \eqref{eq:Z0}. This concludes the proof. \QEDA
\end{pf}
\begin{remark}
The observability condition in Proposition~\ref{prop:nec} can be always fulfilled up to reducing the size of the controller state. On the other hand, later we show in Proposition~\ref{th:prop2} that this property can be automatically enforced.
\end{remark}
\begin{remark}
Proposition~\ref{prop:nec} can be interpreted as an alternative formulation of the wash-out controllability conditions introduced in \cite{1383925,takimoto2007reduced}. These works derive transfer-function-based criteria that guarantee that the controller has a steady-state blocking zero. Moreover, in \cite{takimoto2007reduced} a reduced-order dynamic controller is designed. In the present paper we adopt the structure proposed in \cite{takimoto2007reduced}, as the template for our controller. 
\end{remark}
We are now ready to state the main result of this section, which provides necessary and sufficient condition for the solution to Problem~\ref{prob:state_feedback}. 
\begin{theorem}
Let $\mathcal{H}_0$ be the hybrid system defined as \eqref{eq:HCL} with $d=0$. Assume that $(L, K)$ is observable. Then, $(L, K, G, R)$ solves Problem~\ref{prob:state_feedback} if and only if
\begin{itemize}
    \item[$(i)$] \eqref{eq:Z0} holds
    \item[$(ii)$] the set 
    \begin{equation}
    \label{eq:A0}
    \mathcal{A}_0\coloneqq\{0\} \times [0, T_2]\subset\R^{n+1}
    \end{equation}
    is GES for $\mathcal{H}_0$. 
\end{itemize}
\label{thm:main}
\end{theorem}
\begin{pf}\textit{Proof of necessity}.
The proof of necessity is trivial. Indeed, if $(L, K, G, R)$ solves Problem~\ref{prob:state_feedback}, from Proposition~\ref{prop:nec} it follows that \eqref{eq:Z0} holds. Now, let $d=0$. Then, since $(L, K, G, R)$ solves Problem~\ref{prob:state_feedback}, there exists a compact set 
$$
\mathcal{Q}\subset\{0\}\times \R^{n_c}\times \{0\}\times [0, T_2]
$$
that is GES for $\mathcal{H}_0$. Moreover, by following the same rationals as is in the proof of Proposition~\ref{prop:nec}, it turns out that for all $\phi\in\mathcal{S}_{\mathcal{H}_0}$, 
$\prj_{z}(\Omega(\phi))=\{(0, 0, 0)\}$. These two properties can be used to show that the set $\mathcal{A}_0\subset\mathcal{Q}$ is globally uniformly attractive for $\mathcal{H}_0$. This, due to the homoegenity properties of the flow and jumps maps of $\mathcal{H}_0$, thanks to \cite[Proposition 1]{teel2012lyapunov} enables to conclude that $\mathcal{A}_0$ is GES for $\mathcal{H}_0$.

\textit{Proof of sufficiency}. 
First we show that conditions $(i)$, $(ii)$ imply that $A$ is nonsingular. Assume by contradiction that $A$ is singular i.e. there exist $\overline{x}\neq 0$ such that $A\overline{x}=0$. Select $\overline{\xi}$ such that
$$   
\begin{aligned}
&G\overline{\xi}=(I_{n_c}-L)\overline{\xi}\\
&R\overline{x}=-K\overline{\xi}  
\end{aligned}
$$
this is possible thanks to $(i)$. Notice that by construction 
$$
    \{ \overline{x}, \overline{\xi}, 0\}\in \ker(\mathrm{F}) \cap \ker(I_n-\mathrm{J})
$$
therefore for all $\phi \in \mathcal{S}_{\mathcal{H}_0}(\{(\overline{x},\overline{\xi},0)\}\times[0,T_2])$
$$
\prj_{z}(\Omega(\phi))=\{ \overline{x}, \overline{\xi}, 0\}
$$
which contradicts $(ii)$. We now complete the proof. Let $d\in\R^{\np}$, select $\bar{\xc}\in\R^{\nc}$ such that
$$
\begin{aligned}
&(L-I)\bar{\xc}=GA^{-1}d\\  
&K\bar{\xc}=RA^{-1}d.
\end{aligned}
$$
This is possible due to the satisfaction of \eqref{eq:Z0}. Define 
$\tilde{z}=(\xp-A^{-1}d, \xc-\bar{\xc}, u)$. Then, the dynamics of $\tilde{x}\coloneqq(\tilde{z}, \tau)$ reads
$$
\begin{array}{lll}
&\begin{cases}
\dot{\tilde{z}}=\mathrm{F} \tilde{z}\\
\dot{\tau}=1
\end{cases}&\mathcal{C}\coloneqq\{\tilde{x}\in\xSpace\colon \tau\in[0, T_2]\}\\
&\begin{cases}
\tilde{z}^+=\mathrm{J}\tilde{z}\\
\tau^+=0
\end{cases}&\mathcal{D}\coloneqq\{\tilde{x}\in\xSpace\colon \tau\in[T_1, T_2]\}\\
 \end{array}
$$
which corresponds to the dynamics of $\mathcal{H}_0$. Therefore, since by assumption the set $\mathcal{A}_0$ is GES for $\mathcal{H}_0$, it follows that the set 
$$
\mathcal{A}_d\coloneqq \{-A^{-1}d\}\times\{0\}\times\{0\}\times [0, T_2]
$$
is GES for the closed-loop sytem \eqref{eq:HCL}. This concludes the proof.\QEDA
\end{pf}
\begin{remark}
The fact that the satisfaction of conditions $(i)$, $(ii)$ of Theroem~\ref{thm:main} implies that $A$ is nonsingular makes it possible to solve Problem~\ref{prob:state_feedback} without the need of checking the nonsingularity of the uncertain matrix $A$.
\end{remark}
\section{Controller Design}
In this section we rely on the above results to define a constructive procedure for the state-feedback controller. We chose the controller parameters and define sufficient conditions that guarantee that the controller is a solution to Problem \ref{prob:state_feedback}.
\label{sec:ControllerDesign}
\begin{proposition}
\label{th:prop2}
Suppose that there exist $P\colon[0, T_2]\to \mathbb{S}^{\np+2\nv}$, $\Lambda\in\R^{\nv\times \nv}$, and $\Pi\in\R^{\nv\times \np}$ such that
\begin{align}
\label{eq:flow_LMI}
&\begin{bmatrix}
\dot{P}(\tau)+\He(P(\tau)\mathrm{F}_0) &P(\tau)\mathrm{D}&\mathrm{E}^\top\\
\star&-I&0\\
\star&\star&-I
\end{bmatrix}\prec 0\quad\forall\tau\in[0, T_2]\\
\label{eq:jump_LMI}
&\overline{\mathrm{J}}^\top P(0)\underbrace{\begin{bmatrix}
I&0&0\\
\Pi&\Lambda&0\\
\Pi&\Lambda-I&0
\end{bmatrix}}_{\overline{\mathrm{J}}}-P(\tau)\prec 0,\quad  \forall\tau\in[T_1, T_2]\\
\label{eq:P0_pos}
&P(0)\succ0 
\end{align}
where 
$$
\mathrm{F}_0\coloneqq\begin{bmatrix}
    A_0&0&B_0\\
    0&0&0\\
    0&0&0
\end{bmatrix}, \mathrm{D}\coloneqq\begin{bmatrix}
    D_{\Delta}\\
    0\\
    0
\end{bmatrix},\mathrm{E}\coloneqq\begin{bmatrix}
    E_{\Delta}&F_{\Delta}&0
\end{bmatrix}.
$$
Select $G=R=\Pi$, $L=\Lambda$, and $K=-I+\Lambda$. Then, controller \eqref{eq:Controller} solves Problem~\ref{prob:state_feedback}.
\end{proposition}
\begin{pf}
The proof relies on Theorem~\ref{thm:main}. To this end, first we show that with the proposed selection of the controller parameters, item $(ii)$ in Theorem~\ref{thm:main} holds. By Schur complment from \eqref{eq:flow_LMI}, it follows that for all $\tau\in[0, T_2]$
\begin{equation}
\label{eq:Q_tau}
    Q(\tau)\coloneqq \dot{P}(\tau)+\He(P(\tau)\mathrm{F}_0)+P(\tau)\mathrm{D}\mathrm{D}^\top P(\tau)+\mathrm{E}^\top \mathrm{E}\prec 0
\end{equation}
which implies that
\begin{equation}
\label{eq:P_dot_LMI}
   \dot{P}(\tau)+\He(P(\tau)\mathrm{F}_0)\prec 0 \quad \forall\tau\in[0, T_2]. 
\end{equation}
Therefore, thanks to \eqref{eq:jump_LMI}, \eqref{eq:P0_pos} and \eqref{eq:P_dot_LMI}, by following the same steps as in the proof of \cite[Theorem 2.1]{BRIAT20133449} it turns out that
\begin{equation}
\label{eq:P_pos}
    P(\tau)\succ 0 \quad \forall \tau \in [0,T_2].
\end{equation}
Define for all $\tilde{x}\in\xSpace$
$$
V(\tilde{x})\coloneqq \tilde{z}^\top P(\tau)\tilde{z}.
$$
Thanks to \eqref{eq:P_pos} and the continuity of $P$, there exist $c_1,c_2>0$ such that for all $\tilde{x}\in\xSpace$, 
$$
c_1|\Tilde{x}|^2_{\mathcal{A}_0}\le V(\Tilde{x})\le c_2|\Tilde{x}|^2_{\mathcal{A}_0}
$$
where $\mathcal{A}_0$ is defined in \eqref{eq:A0} and the following identity is used
$$
|\Tilde{x}|_{\mathcal{A}_0}=|\Tilde{z}|\quad \forall \tilde{x}\in \mathcal{C}.
$$
Simple computations reveal that, for all $\tilde{z}\in\mathcal{C}$,
$$
\dot{V}(\tilde{x})\coloneqq\langle\nabla V(\tilde{x}), (\mathrm{F}\tilde{z},1)\rangle=\tilde{z}^\top(\dot{P}(\tau)+\He(P(\tau)\mathrm{F})\tilde{z}.
$$
Hence, thanks to Assumption~\ref{assu: Incertezza}, for some $\Delta\in\mathcal{U}$, one has, for all $\tilde{z}\in\mathcal{C}$, 
$$
\dot{V}(\tilde{x})=\tilde{z}^\top(\dot{P}(\tau)+\He(P(\tau)(\mathrm{F}_0+\mathrm{D}\Delta\mathrm{E}))\tilde{z}.
$$
Thus, by using the following elementary inequality
$$
\He(P(\tau)\mathrm{D}\Delta\mathrm{E})\preceq P(\tau)\mathrm{D}\mathrm{D}^\top P(\tau)+\mathrm{E}^\top\mathrm{E}, \quad\forall \Delta \in \mathcal{U}
$$
thanks to \eqref{eq:Q_tau}, for all $\tilde{x}\in\mathcal{C}$, it holds that
$$
\dot{V}(\tilde{x})\leq \tilde{z}^\top Q(\tau)\tilde{z}.
$$
Since $\tau\mapsto Q(\tau)$ is continuous and $\tau$ belongs to the compact interval $[0, T_2]$, there exists $c\in\R_{\geq 0}$ such that 
\begin{equation}
\label{eq:dot_v_bound}
    \dot{V}(\tilde{x})\leq -c_3 \tilde{z}^\top\tilde{z}\quad \forall \tilde{x}\in\mathcal{C}.
\end{equation}
Let $\Tilde{x} \in \mathcal{D}$. Then
$$
V(J\Tilde{z},0)-V(\Tilde{z},\tilde{\tau}) = \tilde{z}^{\top}(\overline{\mathrm{J}}^{\top} P(0)\overline{\mathrm{J}}-P(\tau))\tilde{z}\prec 0.
$$
Therefore, by using \eqref{eq:jump_LMI}, there exists $c_4>0$ such that
\begin{equation}
\label{eq:dot_v_bound_jump}
    V(J\Tilde{z},0)-V(\Tilde{z},\tilde{\tau})\le -c_4\tilde{z}^{\top}\Tilde{z} \quad \forall \tilde{x}\in \mathcal{D}.
\end{equation}
By combining \eqref{eq:dot_v_bound} and \eqref{eq:dot_v_bound_jump}, from \cite[Theorem 1]{teel2012lyapunov} it follows that the set $\mathcal{A}_0$ is GES for hybrid system $\mathcal{H}_0$. This proves that item $(ii)$ of Theorem \ref{thm:main} holds. To conclude the proof, we show that item $(i)$ holds. From the selection of the controller parameters 
$$
\rank\begin{bmatrix}
    R&K\\G&L-I
\end{bmatrix}=\rank\begin{bmatrix}
    \Pi&\Lambda-I\\ \Pi&\Lambda-I
\end{bmatrix}=\rank\begin{bmatrix}
    \Pi&\Lambda-I
\end{bmatrix}
$$
therefore \eqref{eq:Z0} holds if $\Lambda -I$ is nonsingular. By contradiction assume that $\Lambda -I$ is singular. Let $0 \neq \xi^* \in \ker (\Lambda-I)$. Pick $\phi \in \mathcal{S}_{\mathcal{H}_0}((\underbrace{0,\xi^*,0,}_{\tilde{z}^*}T_2))$. Then since 
$$
\begin{cases}
    &\mathrm{F}\tilde{z}^* = 0\\
    &\mathrm{J}\tilde{z}^* = \tilde{z}^*
\end{cases}
$$
it turns out that for all $(t,j) \in \dom \phi,\quad|\phi(t,j)|_{\mathcal{A}_0}=|\xi^*|\neq0$. This contraddicts the satisfaction of item $(ii)$, thereby showing that $\Lambda-I$ is nonsingular. That implies that item $(i)$ holds. Moreover, since $\Lambda-I$ is nonsingular, the pair $(L,K)=(\Lambda,\Lambda-I)$ is observable. In conclusion all the conditions of Theorem~\ref{thm:main} are fulfilled. This concludes the proof.
\QEDA
\end{pf}
\begin{remark}
The conditions derived in this section focus on guaranteeing stability. They can be extended to include quantitative performance objectives such as exponential decay rate and $\mathcal{H}_\infty$ or $\mathcal{L}_2$‐gain bounds.
\end{remark}
Condition \eqref{eq:jump_LMI} is nonlinear in the decision variables. Hence, it cannot be exploited directly for controller design. To overcome this shortcoming, next we provide sufficient conditions in the form of differential linear matrix inequalities constraints. The approach we use is reminiscent of \cite{daafouz2001parameter} and enable to decouple the (constant) controller parameters from the (clock-dependent) matrix $P$.
\begin{proposition}
\label{th:prop3}
Suppose that there exist $Y \in \R^{n_u\times (n_p+n_u)}$, $S_{11}\in \R^{(n_p+n_u)\times (n_p+n_u)}$, $S_{21} \in \R^{n_u\times(n_p+n_u)}$, $S_{22}\in\R^{n_u\times n_u}$, $W\colon[0, T_2]\to \mathbb{S}^{\np+2\nv}$ such that
\begin{equation}
\label{eq:jump_design}
\begin{bmatrix}
-W(0)&\mathrm{J}_0S +\mathrm{B}_J\begin{bmatrix}
    Y& 0
\end{bmatrix}\\\star&W(\tau)-\He(S)
\end{bmatrix}\prec 0,\quad \tau\in[T_1, T_2],
\end{equation}
and
\begin{equation}
\label{eq:flow_design}
\begin{bmatrix}
-\dot{W}(\tau)+\He(\mathrm{F}_0W(\tau))&\mathrm{D}&W(\tau)\mathrm{
E^\top}\\\star&-I&0\\\star&\star&-I
\end{bmatrix}\prec 0\quad \tau\in[0, T_2]
\end{equation}
where $S\coloneqq \begin{bmatrix}
    S_{11}&0\\S_{21}&S_{22}
\end{bmatrix}$, $\mathrm{J}_0\coloneqq \begin{bmatrix}
    I&0&0\\
    0&0&0\\
    0&-I&0
\end{bmatrix}$, $\mathrm{B}_J\coloneqq \begin{bmatrix}
    0\\
    I\\
    I
\end{bmatrix}$. Select 
$$
\Pi=YS_{11}^{-1}\begin{bmatrix}
I_{\np}\\0_{\nv\times \np}
\end{bmatrix}, \Lambda=YS_{11}^{-1}\begin{bmatrix}0_{\np\times\nv}\\I_{\nv}
\end{bmatrix}.$$ 
Then, controller \eqref{eq:Controller} solves Problem~\ref{prob:state_feedback}.
\end{proposition}
\begin{pf}
We show that the satisfaction of \eqref{eq:jump_design}-\eqref{eq:flow_design} implies the satisfaction of \eqref{eq:flow_LMI}-\eqref{eq:jump_LMI}-\eqref{eq:P0_pos} with $P(\tau) = W^{-1}(\tau)$. From the selection of the controller parameters and the structure of the matrix $S$, one has that
$$
\begin{bmatrix}
\Pi&\Lambda&0  
\end{bmatrix}S=\begin{bmatrix}
    Y&0
\end{bmatrix}.
$$
Therefore
$$
\mathrm{J}_0S +\mathrm{B}_J\begin{bmatrix}
    Y& 0
\end{bmatrix}=\mathrm{J}_0S +\mathrm{B}_J\begin{bmatrix}
    \Pi &\Lambda&0
\end{bmatrix}S = \overline{\mathrm{J}}S.
$$
Thus, the satisfaction of \eqref{eq:jump_design} yields 
$$
\Psi(\tau)\coloneqq\begin{bmatrix}
-P^{-1}(0)&\overline{\mathrm{J}}S\\\star&P^{-1}(\tau)-\He(S)
\end{bmatrix}\prec 0\quad\forall\tau\in[T_1, T_2].
$$
From the former inequality, by congruence one gets, for all $\tau\in[T_1, T_2]$,
\begin{align}
\label{eq:jump_LMI2}
 &\begin{bmatrix}
I\\\overline{\mathrm{J}}^\top
\end{bmatrix}^\top\Psi(\tau)\begin{bmatrix}
I\\\overline{\mathrm{J}}^\top
\end{bmatrix}=-P^{-1}(0)+\overline{\mathrm{J}}P^{-1}(\tau)\overline{\mathrm{J}}^\top\prec 0\\
\label{eq:P0_pos2}
&\begin{bmatrix}
    I\\0
\end{bmatrix}^\top\Psi(\tau)\begin{bmatrix}
 I\\0
\end{bmatrix}=-P^{-1}(0)\prec 0.
\end{align}
Inequality \eqref{eq:P0_pos2} shows that \eqref{eq:P0_pos} holds. Moreover, by taking the Schur complement of \eqref{eq:jump_LMI2} yields the satifsaction of \eqref{eq:jump_LMI}.  To conclude the proof, we now show that the satisfaction of \eqref{eq:flow_design} implies \eqref{eq:flow_LMI} with $P(\tau)=W^{-1}(\tau)$.
Notice that, by using the following elementary identity
$$
\dot{P}(\tau)P^{-1}(\tau)=-P(\tau)\dot{P}^{-1}(\tau)
$$
one obtains
$$
\dot{W}(\tau)=\dot{P}^{-1}(\tau)=-P^{-1}(\tau)\dot{P}(\tau)P^{-1}(\tau).
$$ 
Hence, from \eqref{eq:flow_design}, one has
$$
\begin{bmatrix}
P^{-1}(\tau)\dot{P}(\tau)P^{-1}(\tau)+\He(\mathrm{F}_0P^{-1}(\tau))&\mathrm{D}&P^{-1}(\tau)\mathrm{
E^\top}\\\star&-I&0\\\star&\star&-I
\end{bmatrix}\prec 0
$$
which by congruence implies that \eqref{eq:flow_LMI} holds. This concludes the proof.\QEDA
\end{pf}
\begin{remark}
The conditions in Proposition \ref{th:prop3} are infinite dimensional linear matrix inequalities since they need to be verified for all $\tau$ in an interval. Consequently, standard linear matrix inequalities (\emph{LMI}s) algorithms are not directly applicable. To handle this issue, similarly as in \cite{BRIAT20133449,landicheff2022continuous}, we select $W$ to be a polyomial matrix and rely on Sum-of-Squares (SOS) programming. This allows to check the conditions in Proposition~\ref{th:prop3} via off-the-shelf software as SOSTOOLS \cite{1184594}. To include the constraints on $\tau$ appearing in \eqref{eq:jump_design}-\eqref{eq:flow_design}, we proceed in a similar way as in \cite[Proposition~1 ]{landicheff2022continuous}. An alternative approach to handle conditions such as \eqref{eq:jump_design}-\eqref{eq:flow_design} can be also found in \cite{GEROMEL2019289}. 
\end{remark}
\section{Numerical example}
\label{sec:Examples}
\begin{figure}[t]
    \psfrag{t}{$t$}
    \includegraphics[width=1\columnwidth, trim=0 0 0 0,clip]{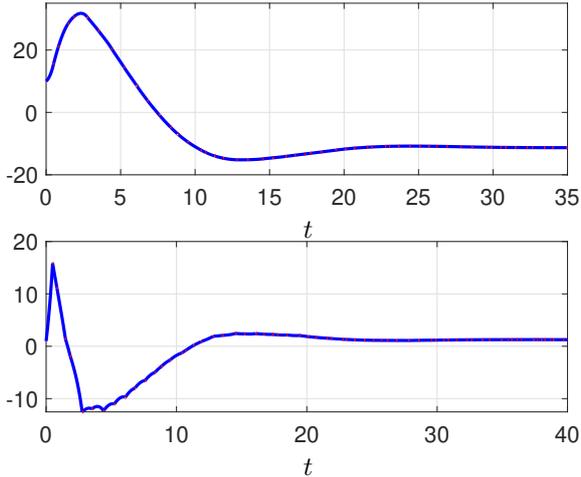}
    \caption{Evolution of the plant state.}
    \vspace{0.5em}
    \label{fig:plat_state}
\end{figure}

\begin{figure}[t]
    \psfrag{t}{$t$}{\includegraphics[width=1\columnwidth,trim=0 0 0 0,clip]{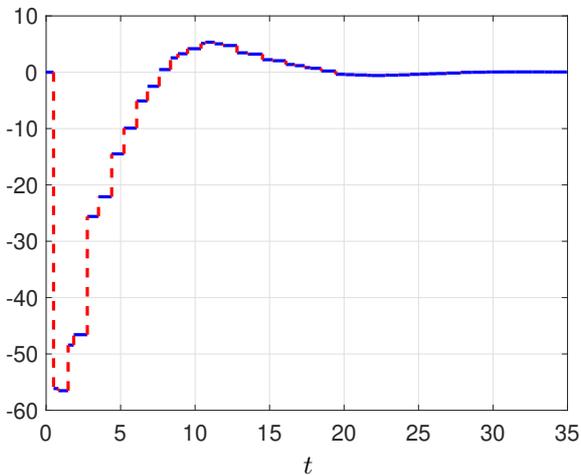}}
    \caption{Evolution of the control input $u$.}
    \label{fig:control}
    \vspace{0.5em}
\end{figure}

We consider the following parameters for plant \eqref{eq:plant} 
$$
\begin{aligned}
    &A_0 = \begin{bmatrix}
        0&1\\1&1
    \end{bmatrix},\quad B_0 = \begin{bmatrix}
        0\\1
    \end{bmatrix}, \quad D = \begin{bmatrix}
        1\\0
    \end{bmatrix},\\  &F = 0.02, \quad E = \begin{bmatrix}
        0.2 & 0
    \end{bmatrix}
\end{aligned}
$$
We solve the conditions in Proposition~\ref{th:prop3} using SOSTOOLS \cite{1184594} and the solver SeDuMi \cite{sturm1999using} by selecting $W$ to be a polynomial matrix of degree $g$. Table~\ref{tab:t2_max_deg} shows how the maximum allowable value of $T_2$ changes according to $g$, when $T_1=0.1$. By selecting $T_1=0.5$ and $T_2=1$, it turns out that the conditions in Proposition~\ref{th:prop3} are feasible with $g=4$. In this case, one gets
$$
\Lambda = 1.0521,\quad \Pi=\begin{bmatrix}
    -1.3830&-2.1917
\end{bmatrix}.
$$
Fig.~\ref{fig:plat_state} and Fig.~\ref{fig:control} report the evolution of the plant state and control input from the initial condition\footnote{Simulations of hybrid systems are performed in MATLAB via the \emph{Hybrid Equations Toolbox} (HyEQ) \cite{sanfelice2013toolbox}.}$\xp(0,0)=(10, 1), \xc(0,0)=0, q(0,0)=0$. Simulations are performed with $\Delta=1$ and $d=(1, 10)$. As expected, simulations  show that the plant state converges to the unforced equilibrium point
$\overline{x}=(-11.25,1.25)$ and the control output converges to zero. 
\begin{table}[htbp]
    \centering
    \caption{Maximum allowable value of $T_2$ versus the degree $g$ of the polynomial matrix $W$ for $T_1 = 0.1$.}
    \label{tab:t2_max_deg}
    \begin{tabular}{@{}S[table-format = 1.0] S@{}}
        \toprule
        {Degree $g$} & {$T_2$} \\
        \midrule
        1 & 0.27 \\
        2 & 0.56 \\
        3 & 0.95 \\
        4 & 1.20 \\
        5 & 1.37 \\
        \bottomrule
    \end{tabular}
\end{table}
\section{Conclusion}
In this paper we addressed the problem of aperiodic sampled-data global exponential stabilization of unknown equilibrium points for uncertain affine plants via. The closed-loop system was modeled as a hybrid dynamical system and incorporated a dynamic state-feedback sampled-data controller. We provided necessary and sufficient conditions for the solution to this problem. The conditions are then used to define a constructive algorithm for controller design exploiting sum-of-squares programming. The effectiveness of the methodology was validated on a numerical example. Future work includes the extension to the case of output feedback control design.
\bibliography{ifacconf}
\appendix
\section{}
\label{App:A}
\textbf{Proof of Lemma~\ref{lemm:singleton}.} By assumption, there exists a compact nonempty set $\mathcal{A}_d$ such that 
$$
\lim_{t+j\to\infty}\vert\phi(t,j)\vert_{\mathcal{A}_d}=0.
$$ 
Therefore, $\Omega(\phi)\subset\mathcal{A}_d$. Hence, in light of Fact~\ref{fact:A_nonsingular}, $A$ is nonsingular and 
\begin{equation}
\label{eq:OmegaLimStruc}
\Omega(\phi)\subset\{-A^{-1}d\}\times\R^{n_c}\times \{0\}\times [0, T_2].  
\end{equation}
Now observe that, since $\mathcal{A}_d$ is compact, it follows that $\phi$ is precompact\footnote{A solution is said to be precompact if it is bounded and complete.}. Thus, since hybrid system \eqref{eq:HybridPlant} satisfies the hybrid basic conditions (see Fact~\ref{fact:HybridBasic}), from \cite[Proposition 6.21, page 129]{Goebel2012}, $\Omega(\phi)$ is forward invariant for \eqref{eq:HCL}. Hence, for all $\chi\in\mathcal{S}(\Omega(\phi))$, thanks to \eqref{eq:OmegaLimStruc} and Assumption~\ref{assu:Bcol}, one has, for all $(t, j)\in\dom\chi$
\begin{equation}
\label{eq:OmegaConstraint}
\begin{aligned}
&\chi_{q}(t, j)=0\\
&R\chi_{\xp}(t, j)+K\chi_\xi(t, j)=0,
\end{aligned}
\end{equation}
where $\chi_{\xp}$, $\chi_\xc$, $\chi_{q}$ indicate, respectively, the $\xp$, $\xc$, and $q$ components of the solution $\chi$. Pick $\overline{x}=(\overline{x}_p, \overline{\xc}, \overline{q}, \overline{\tau})\in\Omega(\phi)$. From \eqref{eq:OmegaLimStruc}, it follows that
$$
\overline{x}_p=-A^{-1}d,\quad \overline{q}=0.
$$
Let $\varphi\in\mathcal{S}(x_1)$. From \eqref{eq:OmegaLimStruc} and \eqref{eq:OmegaConstraint}, one has, for all $(t, j)\in\dom\varphi$
\begin{equation}
\label{eq:identies}
\begin{aligned}
&\varphi_{q}(t, j)=0\\
&\varphi_{\xp}(t, j)=-A^{-1}d\\
&-RA^{-1}d+K\varphi_\xi(t, j)=0\\
\end{aligned}
\end{equation}
Therefore, from the definition of the dynamics of $\xc$, one has for all $(t, j)\in \dom\varphi$
$$
\underbrace{\begin{bmatrix}
K \\
KL\\
\vdots \\
KL^{\nc-1}
\end{bmatrix}}_{O}\varphi_\xi(t,j)=\underbrace{\begin{bmatrix}
-RA^{-1}d\\
(KLG-R)A^{-1}d\\
\vdots\\
(KL^{n-2}G-R)A^{-1}d\\
\end{bmatrix}}_{q},
$$
which implies that
$$
O(\varphi_\xi(t,j)-\varphi_\xi(0,0))=0\quad \forall (t, j)\in \dom\varphi,
$$
Since by assumption $(L,K)$ is observable, matrix $O$ is full column rank. Therefore, $\varphi_\xi(t,j)=\varphi_\xi(0,0)=\overline{\xi}$ and $\overline{\xi}$ corresponds to the unique solution to the following system of linear equations
$$
O\overline{\xi}=q,
$$
which does not depend on the specific selection of the solution $\phi$.  
Hence, by taking $\bar{\xc}=\overline{\xi}$, the result is proven.
\QEDA
\end{document}